\documentclass[12pt, twocolumn]{article}
\usepackage{textcomp} 
\usepackage{geometry}                
\usepackage{graphicx}
\usepackage{amssymb}
\usepackage{epstopdf}
\usepackage{sectsty}
\sectionfont{\large}

\begin{document}

\begin{titlepage}
\begin{center}
{\Large \scshape Origin and evolution of cosmic accelerators - the unique discovery potential of an UHE neutrino telescope}\\
\ \\
{\large Astronomy Decadal Survey (2010-2020) Science White Paper}\\

\ \\

{Interested physicists from the IceCube and ANITA Collaborations \\
(Editors: Pisin Chen$^{1} $, K. D. Hoffman$^{2}$)}

\ \\
{\it \small 1. Kavli Institute for Particle Astrophysics and Cosmology, SLAC National Accelerator Laboratory, Stanford, CA 94305 \\
 2.	Department of Physics, University of Maryland, College Park, MD 20742}

\end{center}

\begin{abstract}
One of the most tantalizing questions in astronomy and astrophysics, namely the origin and the evolution of the cosmic accelerators that produce the highest energy cosmic rays (UHECR), may be best addressed through the observation of ultra high energy (UHE) cosmogenic neutrinos. Neutrinos travel from their source undeflected by magnetic fields and unimpeded by interactions with the cosmic microwave background.  At high energies,  neutrinos could be detected in dense, radio frequency (RF) transparent media via the Askaryan effect.  The abundant cold ice covering the geographic South Pole, with its exceptional RF clarity, has been host to several pioneering efforts to develop this approach, including RICE and ANITA.  Building on the expertise gained in these efforts, and the infrastructure developed in the construction of the IceCube optical Cherenkov observatory, a low-cost array of radio frequency antenna stations could be deployed near the Pole to efficiently detect a significant number of UHE neutrinos with degree scale angular resolution within the next decade.  Such an array, if installed in close proximity to IceCube, could allow cross-calibration on a small but invaluable subset of neutrino events detected by both the optical and radio methods.  In addition to providing critical information in the identification of the source of UHECRs, such an observatory could also provide a unique probe of long baseline high energy neutrino interactions unattainable in any man-made neutrino beam.

\end{abstract}
\end{titlepage}

\section{Neutrinos: a unique astronomical messenger}

Our knowledge of the universe is derived from the observation of fundamental particles that act as messengers, providing a window into their origins. However, photons above 30 TeV have horizons that are limited by pair production due to their interactions with the galactic infrared (IR) and cosmic microwave background (CMB) photons. Further, protons are bent by inter- and intra- galactic magnetic fields, making charged particle astronomy only possible at energies above $10^{19}$ eV.  At higher energies, protons interact with CMB photons through the Greisen-Zatsepin-Kuzmin (GZK) ~\cite{1} process:
\begin{eqnarray*}
p+\gamma_{2.7K} \rightarrow \Delta^+ \rightarrow n+  & \pi^+ & \\
& \hookrightarrow & \mu \nu_{\mu} \\
& & \hookrightarrow e \nu_{e} \nu_{\mu} \\
\end{eqnarray*}
thus a large fraction of the cosmic volume is not accessible with charged particles.  Figure 1 compares the propagation distance of protons and photons to the horizons required for the study of some astrophysical objects. Neutrinos, on the other hand, are only weakly interacting and therefore have a mean free path which exceeds the Hubble radius, allowing the possibility that they could reach Earth from deepest space.  
  
The fact that ultra-high energy cosmic rays (UHECR) have been observed, and almost certainly include a significant proton fraction, guarantees the existence of neutrinos at $10^{17}-10^{19}$ eV, as required by standard-model physics (See Fig.~\ref{fig:GZK}). 
In addition to the GZK neutrinos, there may exist additional UHE neutrinos induced by various relic topological defects. All of these scenarios result in an ultra-high energy neutrino spectrum. It is evident that cosmic neutrinos provide a unique and complementary window into some of the deepest mysteries of our universe. 

\begin{figure}[htbp]
 \includegraphics[width=2.7in]{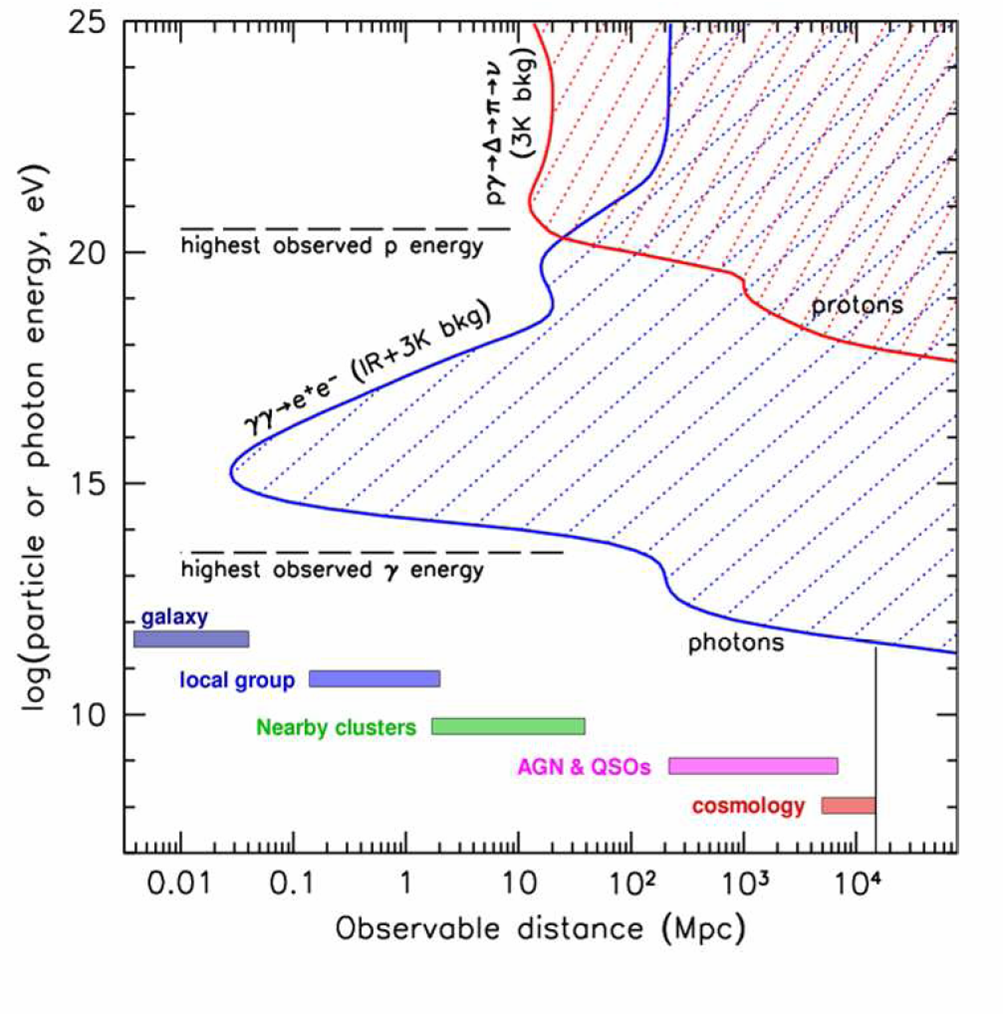}
 \caption{The red (blue) shaded regions show the distances that are inaccessible to protons (photons) as a function of energy. For comparison, the distance scales required for various astrophysical studies are also shown.}
\label{fig:messengers}
\end{figure}

\begin{figure}[htbp]
\includegraphics[width=2.7in]{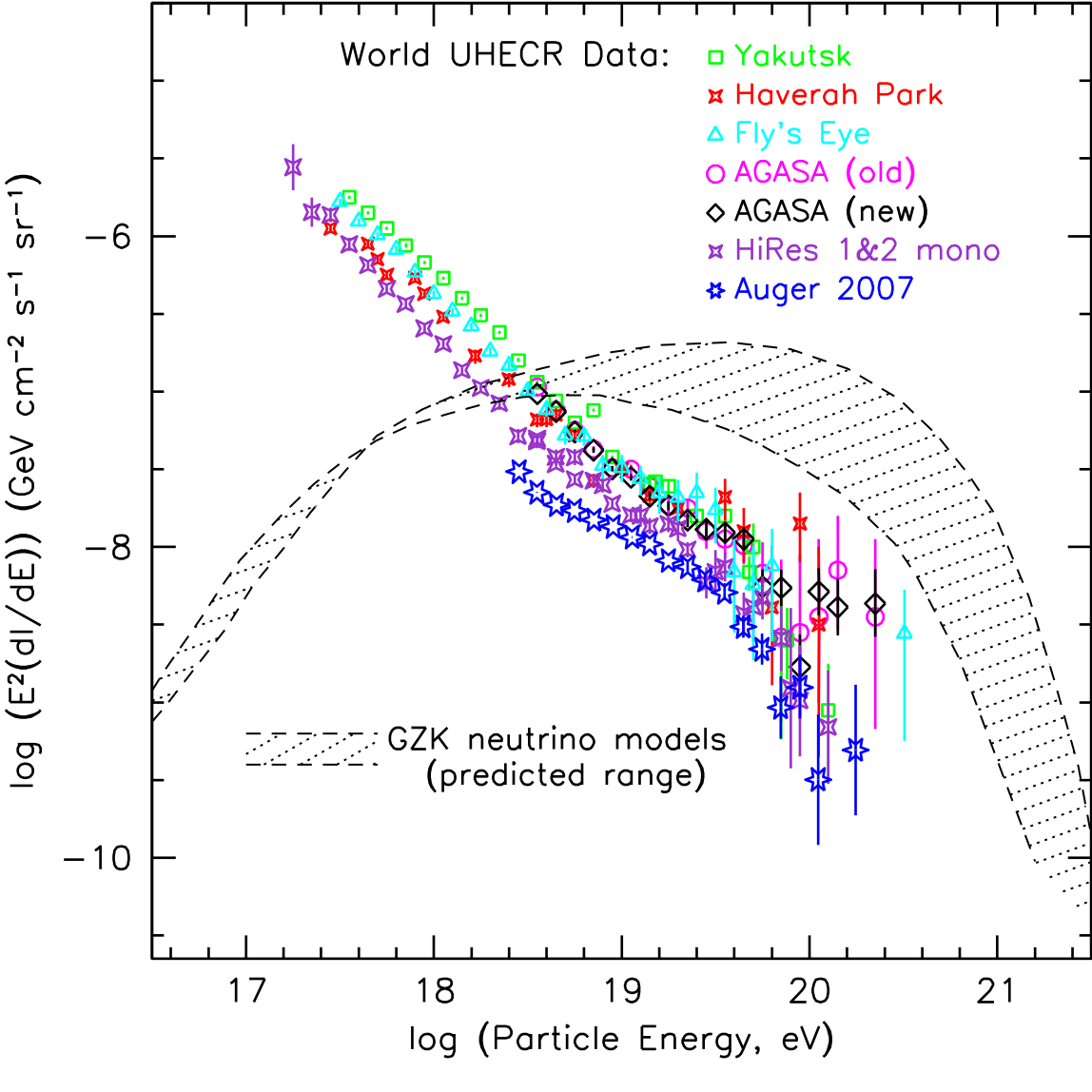}
\caption{World ultra-high energy cosmic ray and predicted cosmogenic neutrino spectrum as of early 2007, including data from the Yakutsk ~\cite{2}, Haverah Park ~\cite{3} the Fly's Eye ~\cite{4},AGASA ~\cite{5}, HiRes ~\cite{6}, and Auger ~\cite{7}, collaborations. GZK neutrino models are from Protheroe \& Johnson ~\cite{8} and Kalashev et al. ~\cite{9}.}
\label{fig:GZK}
\end{figure}

It has been recognized ~\cite{10} that the neutrinos produced in GZK interactions could be used to characterize the UHECR source spectrum and spatial distribution. A study of the UHE neutrino spectrum with good statistics will provide an important piece of the puzzle of the origins of cosmic rays, especially if point sources were resolved. A large and versatile UHE neutrino telescope would also be sensitive to novel aspects of cosmology (such as the Big Bang relic topological defects) and fundamental physics at the energy frontier (such as non-Standard Model neutrino flavor physics and electroweak interactions, Lorentz invariance, extra dimensions, space-time foams, superconducting strings, etc.).  It should also be noted that the study of astrophysical neutrinos in a previously unexplored energy regime could even reveal completely unexpected phenomena. UHE cosmic neutrinos can address a variety of issues in astrophysics and particle physics ~\cite{11},~\cite{12}.  Here we present the scenarios with the greatest discovery potential, including  the issue of the origin and evolution of the ``cosmic accelerators'' that are the most critical to the production of UHECRs, and we propose a path for exploring this regime that could be realized within the next decade. 

\section{Cosmic accelerators}

One of the 11 Science Questions for the New Century put forward by the NRC Turner Committee on ÒConnecting Quarks with the CosmosÓ ~\cite{13} is: ``How do cosmic accelerators work and what are they accelerating?'' Existing models for UHECR can be largely categorized as top-down or bottom-up scenarios. The top-down scenarios assume that the UHECRs are the decay products of some exotic, non-Standard Model particles, which could have been naturally produced at the post-inflationary stage of the universe with masses of $10^{13} - 10^{14}$ GeV. 
The bottom-up scenario, on the other hand, assumes that the UHECRs are ordinary particles, e.g., protons, accelerated at their source to ultra-high energies. 

Observations of the UHECR spectrum above the ankle ($~3\times10^{18}$ eV) from AGASA, HiRes and Auger show both a dip caused by $e^+e^-$ pair production~\cite{17, 18} and a bump consistent with a GZK accumulation clearly visible. In particular, hybrid energy measurements from Auger have reconciled the discrepancy between the HiRes and AGASA energy scales, favoring the HiRes claim of the observation of the GZK cutoff. This evidence supports a simple ansatz where UHECR are accelerated in extragalactic sources~\cite{11} and reaches us over a long baseline, favoring the bottom-up scenario~\cite{19}. 

If the production is indeed bottom-up, what accelerates the cosmic particles and where are the sources? A number of bottom-up cosmic acceleration models have been proposed, where the most developed, diffusive shock acceleration~\cite{14}, could in principle accelerate protons to $10^{21}$ eV. Above these energies, one may have to invoke more exotic models, such as unipolar induction~\cite{15} or plasma wakefield acceleration~\cite{16}. 


Whether the origin of UHECR is top-down or bottom-up and whether their sources are local AGNs, UHE neutrinos are the inescapable by-product of propagation through the cosmic microwave background. Neutrinos will be produced in this GZK process within a distance $(1+z)^{-3}R_{0,GZK}$ from the source of the parent UHECR, where $R_{0,GZK} \sim 50$ Mpc is the local GZK radius. For example, for a UHECR produced by a cosmic accelerator located at $z=1$, the GZK neutrinos would be induced within 6 Mpc from the source. Thus, neutrinos produced by cosmologically distant sources necessarily point back to the source with sub-degree accuracy. This is in distinction to charged particles - it is an open question if UHECR above the GZK cutoff point back to local sources, but old UHECR, which have interacted, and have almost certainly diffused from their source position on the sky.


An EHE neutrino observatory large enough to amass a sample on the order of  hundreds of events  would provide two unique discovery opportunities based on the detection of GZK neutrinos.  First,  the GZK neutrino spectrum and directions would be  indispensable in a multi-particle astronomical analysis to determine  the sources of the highest energy particles in the Universe. Once the sources for UHECRs are identified, it may be possible to further investigate the evolution of the cosmic accelerators through the redshift dependence of their host galaxies.  Complementing the astrophysical implications, mere detection of neutrino induced  events would extend our knowledge of neutrino properties. By measuring the event rate as a function of nadir angle, the opacity of the Earth can be used to determine neutrino-nucleon cross sections at center of mass energies unavailable to any current or planned laboratory facility. For example, a small cross section compared to standard model extrapolations could indicate non-perturbative aspects of nucleon structure, while a cross section increasing in energy could be a doorway to multi-dimensional physics beyond the standard model.


\section{Radio detection of neutrinos}

Ironically, the very lack of interactivity that makes neutrinos so valuable in probing dense objects from cosmological distances presents a major challenge to their detection, even at GZK energies where their cross sections are enhanced.   In addition, at GZK energies, the flux  will be low (see Fig.~\ref{fig:GZK}).  The technique used by the current generation of neutrino observatories relies on the detection of optical Cherenkov radiation principally from secondary leptons or cascades produced in neutrino-nucleon deep inelastic scattering, with large natural reservoirs of water or ice serving as both a target medium and a Cherenkov radiator.  The absorption and scattering of light in these media require a density of instrumentation that is too costly to scale up to an array of the size required for GZK studies. 
An alternate detection mechanism was suggested as early as 1962 when Askaryan~\cite{22} proposed that high energy showers might produce coherent radio emission in dense media. These emissions would arise as an excess of negative charge builds up as electrons are swept out along a relativistically advancing shower front (20\% more electrons than positrons when the shower is fully developed). The longer wavelength components of the broadband radiation from the motion of this large net negative charge will add coherently, while for smaller optical wavelengths the individual particle sources will contribute to the radiation fields with essentially random phases. The net result gives rise to a radio frequency (RF) impulse. The shower dimensions determine the wavelengths over which these emissions are coherent, with the amount of power radiated in radio exceeding the optical for shower energies greater than about $10^{15}$ eV. The ``Askaryan effect'' has been demonstrated in a series of experiments at SLAC by directing pulses of electrons into sand, salt and ice~\cite{23,24,25,26}. The exceptional RF clarity of cold ice suggests that a sparse array of antennas embedded in the South Polar Ice Cap may provide a cost effective approach to the detection of GZK neutrinos.

Any design for a future GZK neutrino observatory will rely heavily on the ongoing pioneering efforts to exploit the Askaryan effect for UHE neutrino detection.  The most ambitious of these is ANITA~\cite{30,31}, a balloon borne antenna array which, for a few weeks in the austral summers of 2006-2007~\cite{32} and 2008-2009, surveyed the entire continent for RF emissions emanating from horizontal neutrino-induced showers that are refracted at the surface of the Antarctic ice. While ANITA may be poised to observe the first GZK neutrino, their synoptic approach is season-limited, making it difficult to collect sufficient statistics for study.  In addition, their vantage point limits their ability to reject surface noise by reconstructing a three dimensional interaction vertex within the volume of the ice.  A complementary approach is taken by RICE~\cite{27}, a small grid of submerged antennas embedded in the South Polar Icecap at depths of 100-300m.  While this approach delivers a decreased effective volume compared to ANITA, it has the unique capability of making the in situ measurements of the RF properties of the ice (including measurements of attenuation length, noise, and birefringence) that will be essential in the design and simulation of any GZK scale array, as well as the demonstrated ability to reconstruct three dimensional vertices.  In addition, RICE has been a testbed for the technical design and analysis techniques that would be employed in any englacial array.

Recently, a unique opportunity has been presented by the ongoing construction of the IceCube neutrino observatory, which, with an instrumented area of a cubic kilometer, is the largest optical Cherenkov observatory in existence. By installing antenna arrays parasitically in holes drilled for IceCubeÕs phototubes and using IceCubeÕs power and communication infrastructure, the RF properties of the South Pole ice are being studied over a longer baseline (up to a kilometer) and at greater depths (down to 1450m) than allowed by RICE.  By marrying the expertise developed for RICE, ANITA, and IceCube, this was done quickly, and at minimal cost.  To date, five clusters of receivers containing four channels each, and six transmitters have been deployed within the IceCube footprint~\cite{aura}.

These deployments suggest an obvious path toward a GZK scale array that exploits the infrastructure at the Pole, the exceptional volume and clarity of the ice there, and a synergy between the well established optical, and the pioneering radio technique.
 At UHE, the secondary leptons that IceCube relies upon for neutrino detection may propagate for 20-30 km or more (in the case of muons or taus) before reaching the array~\cite{33}. This potentially long propagation distance leads to an unknown amount of lost energy.
 The kinematics of the event is such that the lepton typically carries 75-80\% of the primary neutrino energy, with the remainder being deposited into a local hadronic cascade initiated by debris recoiling from the initiating charged current interaction.  The energy deposited in this cascade can only be measured by  IceCube if it is contained within the instrumented volume. However, while initiated by hadrons, the cascade rapidly develops into an $e^+e^-\gamma$  shower in ice, which would be detectable through Askaryan emission. The simultaneous detection of an event with both technologies, therefore, is possible, and would provide complementary information.  
  
\begin{figure}[htbp]
 \includegraphics[width=2.7in]{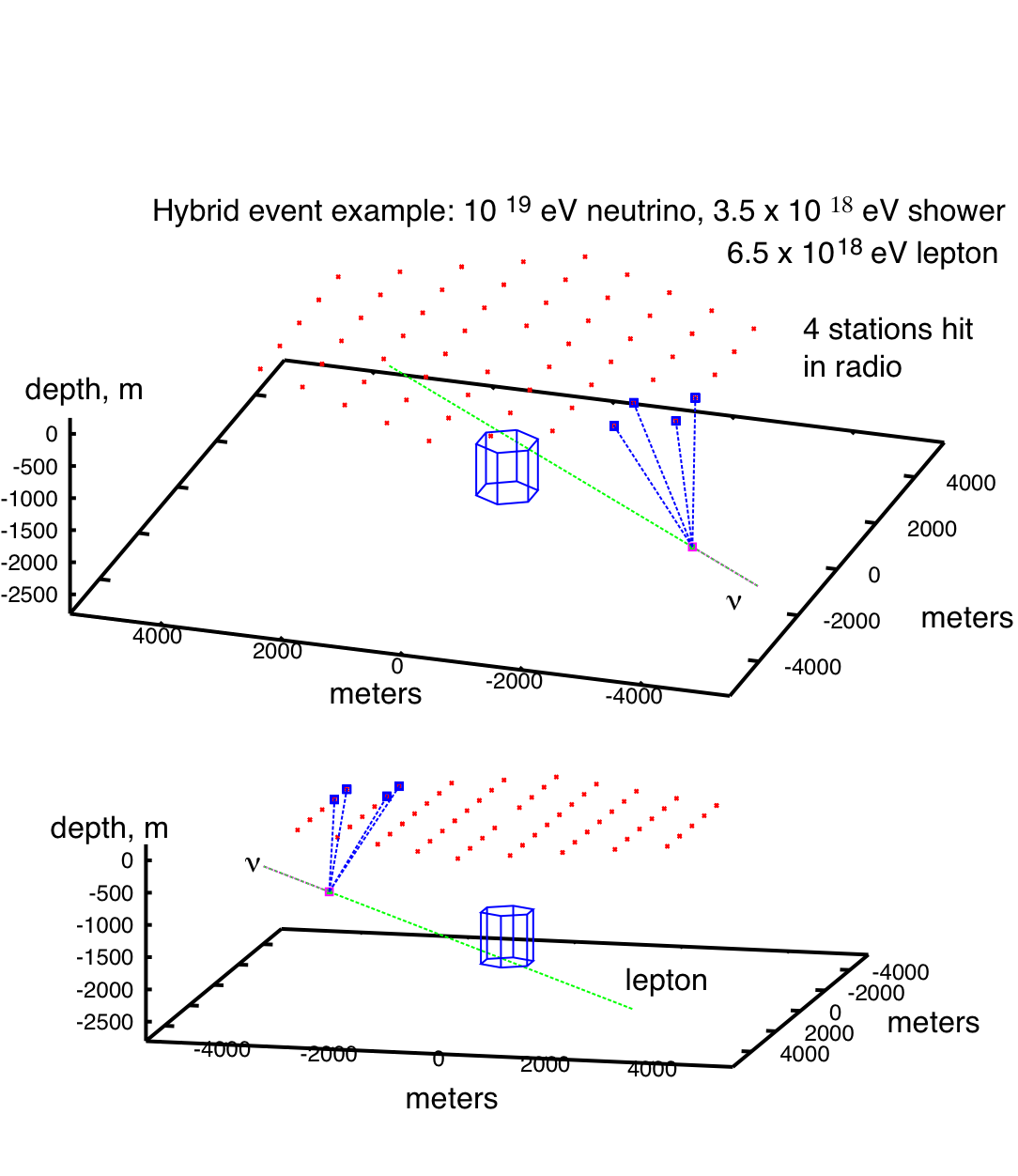}
 \caption{Simulation of an incident $10^{19}$ eV neutrino which deposits 35\% of its energy into a shower that was seen by 4 of the sub-surface radio detectors (red dots), while the secondary lepton passes just outside the IceCube array (blue hexagon) and is detected with an energy of $6.5 \times 10^{18}$ eV. }
\label{fig:IceRay}
\end{figure}

 A concentric  array of RF antennas surrounding IceCube (Fig.~\ref{fig:IceRay}) could observe the radio emission from the primary vertex of some of the same events that produce detectable leptons in IceCube. Even a relatively coarse array with km scale spacing between antenna clusters may detect the strong coherent radio impulses from the cascade vertex.  Although such hybrid events would make up only a small fraction of the event sample, even a few cross calibrated events would provide an invaluable verification of the radio technique.    A parallel investigation into  the detection of acoustic signals initiated by thermal energy deposited by UHE neutrino interactions is ongoing~\cite{acoustic} and may also be deployed, at small incremental cost, in a hybrid detection scenario~\cite{34}.  It should be noted that cold ice is the only environment where all three of these methods could be used simultaneously.

\section{A phased approach to an IceCube radio extension}

As outlined in section 2, it will be necessary to collect $\mathcal{O}(100)$ GZK neutrino events in order to attain adequate statistics to exploit the science potential offered by GZK neutrinos. In our consideration of the event rates, we have to be realistic about the confidence with which we can anticipate the GZK flux. Despite significantly improved data on the spectrum near and beyond the ÒankleÓ, observations can still be accommodated with a wide range of assumptions regarding the injection slope at the source, the cosmological evolution of the sources and the composition at injection. 
Accounting for AUGER observations, current bottom up scenarios yield GZK event
 rates ranging from unobservable (although this requires an unlikely
 conspiracy of parameters; see~\cite{35}) to $\sim 0.1$ per km$^2$ per year.
 Experimental limits from ANITA and RICE allow event rates larger by a
factor of 10, leaving room for enhanced cross-sections or direct
contributions from the acceleration source to the total neutrino flux. To
be prudent, we envision a phased approach to building an RF array, where
a modest initial installation could be expanded after the presence of a
signal was confirmed and the flux measured.

{\bf Phase-I: Prototype Testbed}~\cite{jkelley}  
Continuing the measurements started with RICE and the IceCube codeployments, a prototype antenna station would provide a comprehensive temporal measurement of the detected power spectrum in the 30-to-1000-MHz range down to power levels of -110 dBm/MHz for both continuous and episodic events, the radio signal transmission as a function of the sub-surface antenna depth, and long-term monitoring of RFI backgrounds at the South Pole.

{\bf Phase-II: 50-100 km$^2$ Radio Antenna Array}
As an intermediate step, one may aim at an event rate of 3-5 GZK neutrinos per year, perhaps requiring coverage of an area of 50-100 km$^2$ around IceCube. The guiding principle for the design should be that it would deliver degree-scale precision in the reconstruction of the incoming neutrino directions, which should be sufficient for resolving the locations of the astrophysical sources for cosmic accelerators. 


{\bf Phase-III: Full-Scale Array}
Assuming an event rate of 3-5 GZK neutrinos per year, Phase II should be able to make a definitive statement about UHECR cosmic accelerators assuming 10 years of operation. We note, however, that the antenna array would be extendable. Based on the experience gained and better knowledge on the GZK neutrino flux through Phase-II, the array could be extended to 300-1000 ${\rm km}^2$, capable of detecting at least 20-50 GZK neutrinos per year. This would not only help to expedite the data collection, but also allow the pursuit of additional frontier astrophysics and fundamental physics questions.

\section{Summary}

We point out that one critical question in astronomy and astrophysics, namely the origin and the evolution of the cosmic accelerators that produce the highest energy cosmic rays, may be best addressed through the detection of ultra high energy (UHE) cosmic neutrinos. Using existing technology, we believe that this can be realized using a low-cost array of radio antennas surrounding IceCube. Such a UHE neutrino telescope would extend neutrino astronomy to ExaVolt energies, yielding substantial rates of cosmogenic neutrinos, the so-called GZK or ``guaranteed'' neutrinos, and determine their direction to degree-scale precision, allowing the identification of the sources of the highest energy cosmic rays.  Such an observatory has the added  benefit of providing a probe of neutrino interactions at energies unattainable in the laboratory.  Building such an array near IceCube may also allow the detection of a small number of hybrid events, yielding a cross calibration of the two techniques.

\end{document}